\documentclass[twocolumn]{aastex63}
\usepackage{lineno}

\usepackage[utf8]{inputenc}
\usepackage{natbib}
\usepackage{graphicx}
\usepackage{gensymb}
\usepackage[english]{babel}
\usepackage{wrapfig}
\usepackage{appendix}
\usepackage{float}
\usepackage{booktabs}
\usepackage[para]{threeparttable}

\begin{document}

\title{The Green Bank Northern Celestial Cap Pulsar Survey. VI. Discovery and Timing of PSR J1759+5036: A Double Neutron Star Binary Pulsar}

\author[0000-0001-5134-3925]{G.~Y.~Agazie}
\affiliation{Dept. of Physics and Astronomy, West Virginia University, Morgantown, WV 26501}
\affiliation{Center for Gravitational Waves and Cosmology, West Virginia University, Chestnut Ridge Research Building, Morgantown, WV 26505, USA}
\affiliation{Center for Gravitation, Cosmology, and Astrophysics, 
Dept. of Physics, University of Wisconsin-Milwaukee, P.O. Box 413, Milwaukee, WI 53201, USA}

\author{M.~G.~Mingyar}
\affiliation{Dept. of Physics and Astronomy, West Virginia University, Morgantown, WV 26501}
\affiliation{Center for Gravitational Waves and Cosmology, West Virginia University, Chestnut Ridge Research Building, Morgantown, WV 26505, USA}
\affiliation{Dept. of Physics, Montana State University, Bozeman, MT, 59717, USA}

\author[0000-0001-7697-7422]{M.~A.~McLaughlin}
\affiliation{Dept. of Physics and Astronomy, West Virginia University, Morgantown, WV 26501}
\affiliation{Center for Gravitational Waves and Cosmology, West Virginia University, Chestnut Ridge Research Building, Morgantown, WV 26505, USA}

\author[0000-0002-1075-3837]{J.~K.~Swiggum}
\affiliation{Dept. of Physics, 730 High St., Lafayette College, Easton, PA 18042, USA}

\author[0000-0001-6295-2881]{D.~L.~Kaplan}
\affiliation{Center for Gravitation, Cosmology, and Astrophysics, 
Dept. of Physics, University of Wisconsin-Milwaukee, P.O. Box 413, Milwaukee, WI 53201, USA}

\author[0000-0003-4046-884X]{H.~Blumer}
\affiliation{Dept. of Physics and Astronomy, West Virginia University, Morgantown, WV 26501}
\affiliation{Center for Gravitational Waves and Cosmology, West Virginia University, Chestnut Ridge Research Building, Morgantown, WV 26505, USA}

\author[0000-0002-3426-7606]{P.~Chawla}
\affiliation{Dept.~of Physics and McGill Space Institute, McGill Univ., Montreal, QC H3A 2T8, Canada}

\author[0000-0002-2185-1790]{M.~DeCesar}
\affiliation{Dept. of Physics, 730 High St., Lafayette College, Easton, PA 18042, USA}

\author[0000-0002-6664-965X]{P.~B.~Demorest}
\affiliation{National Radio Astronomy Observatory, P.O. Box O, Socorro, NM
87801, USA}

\author[0000-0001-5645-5336]{W. Fiore}
\affiliation{Dept. of Physics and Astronomy, West Virginia University, Morgantown, WV 26501}
\affiliation{Center for Gravitational Waves and Cosmology, West Virginia University, Chestnut Ridge Research Building, Morgantown, WV 26505, USA}

\author[0000-0001-8384-5049]{E.~Fonseca}
\affiliation{Dept.~of Physics and McGill Space Institute, McGill Univ., Montreal, QC H3A 2T8, Canada}

\author[0000-0003-4679-1058]{J.~D.~Gelfand}
\affiliation{Center for Astro, Particle, and Planetary Physics, New York University Abu Dhabi, PO Box 129188,  Abu Dhabi, UAE}
\affiliation{Affiliated Member, Center for Cosmology and Particle Physics, New York University, New York, NY, 10276, USA}

\author[0000-0001-9345-0307]{V.~M.~Kaspi}
\affiliation{Dept.~of Physics and McGill Space Institute, McGill Univ., Montreal, QC H3A 2T8, Canada}

\author[0000-0001-8864-7471]{V.~I.~Kondratiev}
\affiliation{ASTRON, the Netherlands Institute for Radio Astronomy, Oude Hoogeveensedijk 4, 7991 PD Dwingeloo, The Netherlands}

\author{M.~LaRose}
\affiliation{Dept. of Physics and Astronomy, West Virginia University, Morgantown, WV 26501}
\affiliation{Center for Gravitational Waves and Cosmology, West Virginia University, Chestnut Ridge Research Building, Morgantown, WV 26505, USA}

\author[0000-0001-8503-6958]{J.~van Leeuwen}
\affiliation{ASTRON, the Netherlands Institute for Radio Astronomy, Oude Hoogeveensedijk 4, 7991 PD Dwingeloo, The Netherlands}

\author[0000-0002-2034-2986]{L.~Levin}
\affiliation{Jodrell Bank Centre for Astrophysics, School of Physics and Astronomy, The University of Manchester, Manchester, M13 9PL, UK}

\author[0000-0002-2972-522X]{E.~F.~Lewis}
\affiliation{Dept. of Physics and Astronomy, West Virginia University, Morgantown, WV 26501}
\affiliation{Center for Gravitational Waves and Cosmology, West Virginia University, Chestnut Ridge Research Building, Morgantown, WV 26505, USA}

\author[0000-0001-5229-7430]{R.~S.~Lynch}
\affiliation{Green Bank Observatory, P.O. Box 2, Green Bank, WV 24494, USA}

\author[0000-0001-5481-7559]{A.~E.~McEwen}
\affiliation{Center for Gravitation, Cosmology, and Astrophysics, 
Dept. of Physics, University of Wisconsin-Milwaukee, P.O. Box 413, Milwaukee, WI 53201, USA}

\author[0000-0002-4187-4981]{H.~Al Noori}
\affiliation{Dept. of Physics, University of California, Santa Barbara, CA 93106, USA}

\author[0000-0002-0430-6504]{E.~Parent}
\affiliation{Dept.~of Physics and McGill Space Institute, McGill Univ., Montreal, QC H3A 2T8, Canada}

\author[0000-0001-5799-9714]{S.~M.~Ransom}
\affiliation{National Radio Astronomy Observatory, 520 Edgemont Rd., Charlottesville, VA 22903, USA}

\author{M.~S.~E.~Roberts}
\affiliation{Center for Astro, Particle, and Planetary Physics, New York University Abu Dhabi, PO Box 129188,  Abu Dhabi, UAE}
\affiliation{Eureka Scientific, Inc. 2452 Delmer Street Suite 100
Oakland, CA 94602, USA}

\author[0000-0003-4391-936X]{A.~Schmiedekamp}
\affiliation{Dept. of Physics, The Pennsylvania State University, Ogontz Campus, Abington, Pennsylvania 19001, USA}

\author[0000-0002-1283-2184]{C.~Schmiedekamp}
\affiliation{Dept. of Physics, The Pennsylvania State University, Ogontz Campus, Abington, Pennsylvania 19001, USA}

\author[0000-0002-7778-2990]{X.~Siemens}
\affiliation{Dept. of Physics, Oregon State University, Corvallis, OR 97331, USA}

\author[0000-0002-6730-3298]{R.~Spiewak}
\affiliation{Jodrell Bank Centre for Astrophysics, School of Physics and Astronomy, The University of Manchester, Manchester, M13 9PL, UK}
\affiliation{Centre for Astrophysics and Supercomputing, Swinburne University of Technology, PO Box 218, Hawthorn, VIC 3122, Australia}

\author[0000-0001-9784-8670]{I.~H.~Stairs}
\affiliation{Dept. of Physics and Astronomy, University of British Columbia, 6224 Agricultural Road, Vancouver, BC V6T 1Z1 Canada}

\author[0000-0002-9507-6985]{M. Surnis}
\affiliation{Jodrell Bank Centre for Astrophysics, School of Physics and Astronomy, The University of Manchester, Manchester, M13 9PL, UK}


\begin{abstract}
The Green Bank North Celestial Cap (GBNCC) survey is a 350-MHz  all-sky survey for pulsars and fast radio transients using the Robert C. Byrd Green Bank Telescope. To date, the survey has discovered over 190 pulsars, including 33 millisecond pulsars (MSPs) and 24 rotating radio transients (RRATs). Several exotic pulsars have been discovered in the survey, including PSR~J1759+5036, a binary pulsar with a 176-ms spin period in an orbit with a period of 2.04 days, an eccentricity of 0.3, and a projected semi-major axis of 6.8\,light seconds. Using seven years of timing data, we are able to measure   one post-Keplerian parameter,  advance of periastron, which has allowed us to constrain the total system mass to $2.62 \pm 0.03 \; M_{\odot}$. This constraint, along with the spin period and orbital parameters, suggests that this is a double neutron star system, although we cannot entirely rule out a pulsar-white dwarf binary. This pulsar is only detectable in roughly 45\% of observations, most likely due to scintillation. However,  additional observations are required to determine whether there may be other contributing effects.

\end{abstract}



\section{Introduction}
The Green Bank North Celestial Cap (GBNCC) pulsar survey\footnote{\url{http://astro.phys.wvu.edu/GBNCC/}} is a comprehensive survey of the northern celestial sky ($\delta>-40\degree$) at 350 MHz \citep{Stovall_2014} with the Robert C. Byrd Green Bank Telescope (GBT) in West Virginia. The GBNCC survey is projected to cover $\sim$80\% of the entire sky with  sensitivity to millisecond pulsars (MSPs), canonical pulsars, and sources of isolated dispersed pulses such as rotating radio transients (RRATs) and fast radio bursts (FRBs) \citep{RRAt_discovery,2007_frb_discovery,thornton_frb}. The nominal sensitivity to pulsars is 0.74\,mJy with an 6\% duty cycle, which corresponds to a dispersion measure of 0~pc~cm$^{-3}$, spin period of $\sim 1$~s, and sky temperature of 95~K, as shown in Figure~3 of \citet{McEwen_2020}.

Pulsars with a binary companion comprise about 6\% of all known pulsars \citetext{ATNF Pulsar Catalogue v1.63; \citealp{ATNF_cat}}. Most MSPs are in binary systems, as they are spun-up to millisecond periods through angular momentum transfer from a   companion through Roche Lobe overflow \citep{roche_lobe_orig,binary_living_review}. This accretion process typically results in an MSP-white dwarf system in an extremely circular orbit, with eccentricity of $10^{-5}$ $\lessapprox$ e $ \lessapprox $ 0.01 \citep{Pulsars_as_probes}.

Some binary pulsars take a different evolutionary track. If the companion is massive enough to also undergo a supernova, and the kick of both explosions does not destabilize and break up the system, then a double neutron star (DNS) binary will result \citetext{e.g. \citealp{Tauris_2017}}. In this case, the first-born neutron star will be 'recycled', and the second-born neutron star will be a canonical pulsar. It has been observed that recycled DNS pulsars have spin periods ($P$) ranging from 16.9\,ms to 185\,ms and period derivative ($\dot{P}$) range of $2.2\times 10^{-20}$\,s\,s$^{-1}$ to $1.7\times 10^{-17}$\,s\,s$^{-1}$  \citep{ATNF_cat,Tauris_2017,DNS_merger_rate}. 

Currently, 15 DNS systems are known; in 13 of them, we have detected the recycled pulsar, with $P$ ranging from  16.9\,ms to 185\,ms \citep{DNS_merger_rate}. In two systems, however, the younger canonical pulsar has been detected, with $P$ of 144\,ms and 2.7\,s, and $\dot{P}$ of $2.0\times 10^{-14}$\,s\,s$^{-1}$ and $8.9\times 10^{-16}$\,s\,s$^{-1}$ respectively. The DNS systems have eccentricities ranging from 0.08 to 0.83 \citetext{ATNF Pulsar Catalogue v1.63; \citealp{ATNF_cat}}. Some DNS systems have a measurable post-Keplerian (PK) parameter, which can be used to place constraints on the total system, pulsar and companion masses. Measurement of two PK parameters can fully constrain the pulsar and companion masses and other system parameters such as orbital inclination \citep{dd_model_1,Taylor_1989,Damour_1992}. A third makes the system over-determined, meaning it can be used to test theories of gravity by calculating the predicted value for the third PK parameter given the first two and comparing it to the measured value \citep{stairs_review,kramer_2006}.

In this paper, we report the discovery and properties of PSR~J1759+5036, a 176-ms binary pulsar in a 2-day, eccentric orbit. Its small $\dot{P}$ suggests partial recycling, which is consistent with the properties of DNS systems. We present evidence, including a mass constraint, to suggest this system is likely a DNS system.

\section{Discovery and Timing Observations} 

We first detected PSR~J1759+5036 in single-pulse search output from the discovery observation of \object[PSR J1800+5034]{PSR J1800+5034}, an unrelated 578-ms pulsar with a dispersion measure (DM) of 22.7\,pc\,cm$^{-3}$ published in 2018 \citep{Lynch_2018}.  The single pulses were detected at a DM of 7.77\,pc\,cm$^{-3}$  two years after timing observations ceased for the former source. The pulsar was subsequently detectable in a periodicity search with a spin period of 176\,ms, and timed using the archival data of the PSR~J1800+5034 observation campaign, consisting of 24 observations taken on the GBT from February 2013 to January 2014 at 820\,MHz with time resolution of 81.92\,$\rm \mu s $ and 2048 frequency channels.  Dedispersing at the DM of 7.77\,pc\,cm$^{-3}$ and refolding at the 176-ms spin period of PSR~J1759+5036 produced 10 detections. As PSR~J1759+5036 was not the source of interest for these observations, it was not at the center of the beam when data were being taken, possibly reducing the detected flux. 
 
\begin{figure}
  \includegraphics[width=\linewidth]{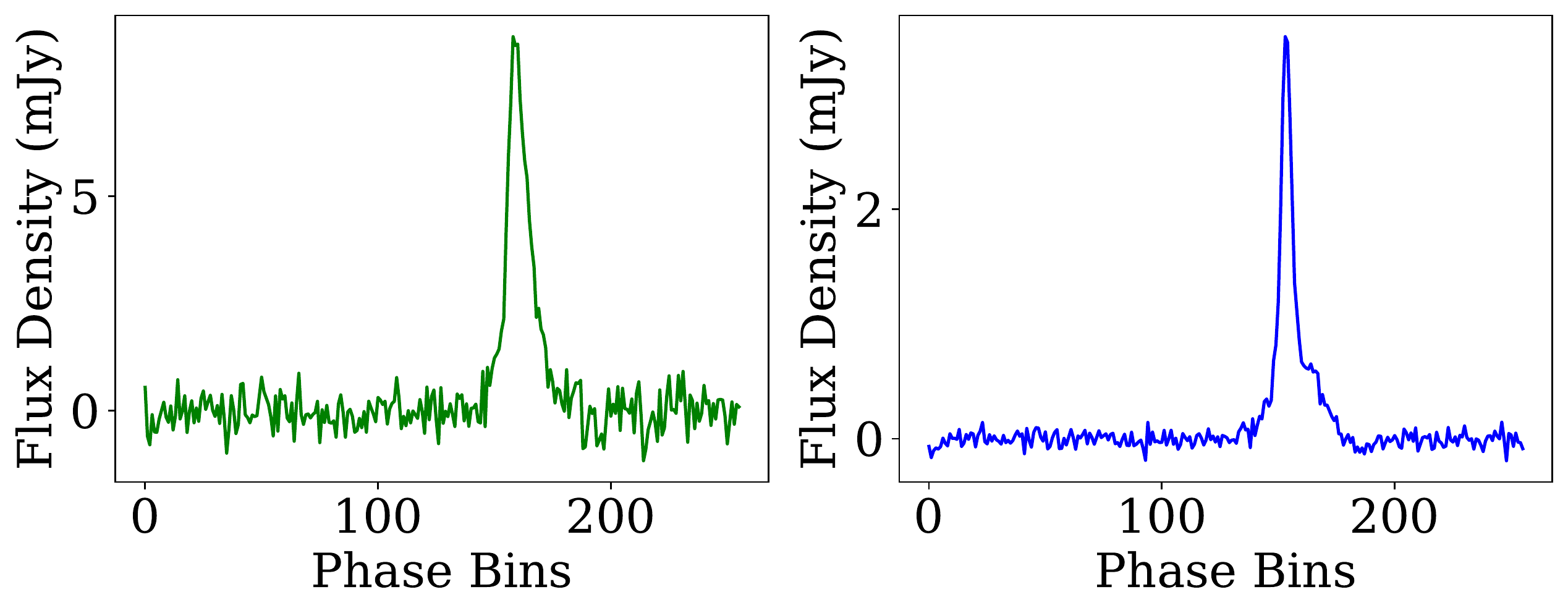}
  \caption{Composite profiles of PSR~J1759+5036 at 350 (left) and 820 (right) MHz with integration times of 4.2\,hr and 6.4\,hr respectively. Both profiles have 256 phase bins and were used as standard profiles to recalculate TOAs in order to refine the timing ephemeris.} 
  \label{fig:profiles}
\end{figure}

The entire timing dataset for PSR~J1759+5036 spans 2013--2020 with 109 total observations of which 50 were detections. Observations at both frequencies ranged from 3 to 60 minutes, A summary of the timeline of observation runs can be found in Table~\ref{tab:observations}. For all observations, the Green Bank Ultimate Pulsar Processing Instrument (GUPPI) backend was used, and radio frequency interference (RFI) was excised using the 
\verb+pazi+ tool in the \verb+PSRCHIVE+ software package\footnote{\url{http://psrchive.sourceforge.net/}}. We then summed each observation in the time and frequency domain to generate best profiles to compute times of arrival (TOAs) using the tool \verb+pat+. For most observations, we calculated 4--7 TOAs, depending on detection signal-to-noise ratio (S/N), with a few marginal detections only producing 1--3 useful TOAs. We made composite profiles for 350 and 820-MHz data (Figure~\ref{fig:profiles}) by adding together all the observations of a particular frequency using the \verb+psradd+ tool. We then  used \verb+paas+ to generate a standard profile which was then used to recalculate TOAs in order to refine the timing ephemeris.

 \begin{table}
     \centering
     \begin{tabular}{c|c|c|c}
          Time frame & Frequency (MHz) & N$_{obs}$ & Detections \\
          \hline
          2013--2014 & 820 & 24 & 10 \\
          2017--2019 & 820/350 &54 & 20 \\
          10/2019 & 350 & 7 & 5 \\
          12/2019 -- 01/2020 & 820 & 24 & 15 \\
     \end{tabular}
     \caption{Summary of observations for PSR~J1759+5036 with the number of observations (N$_{obs}$), observing frequencies, and detections per observation run indicated.}
     \label{tab:observations}
 \end{table}
 
\begin{figure}
    \centering
    \includegraphics[width=\linewidth]{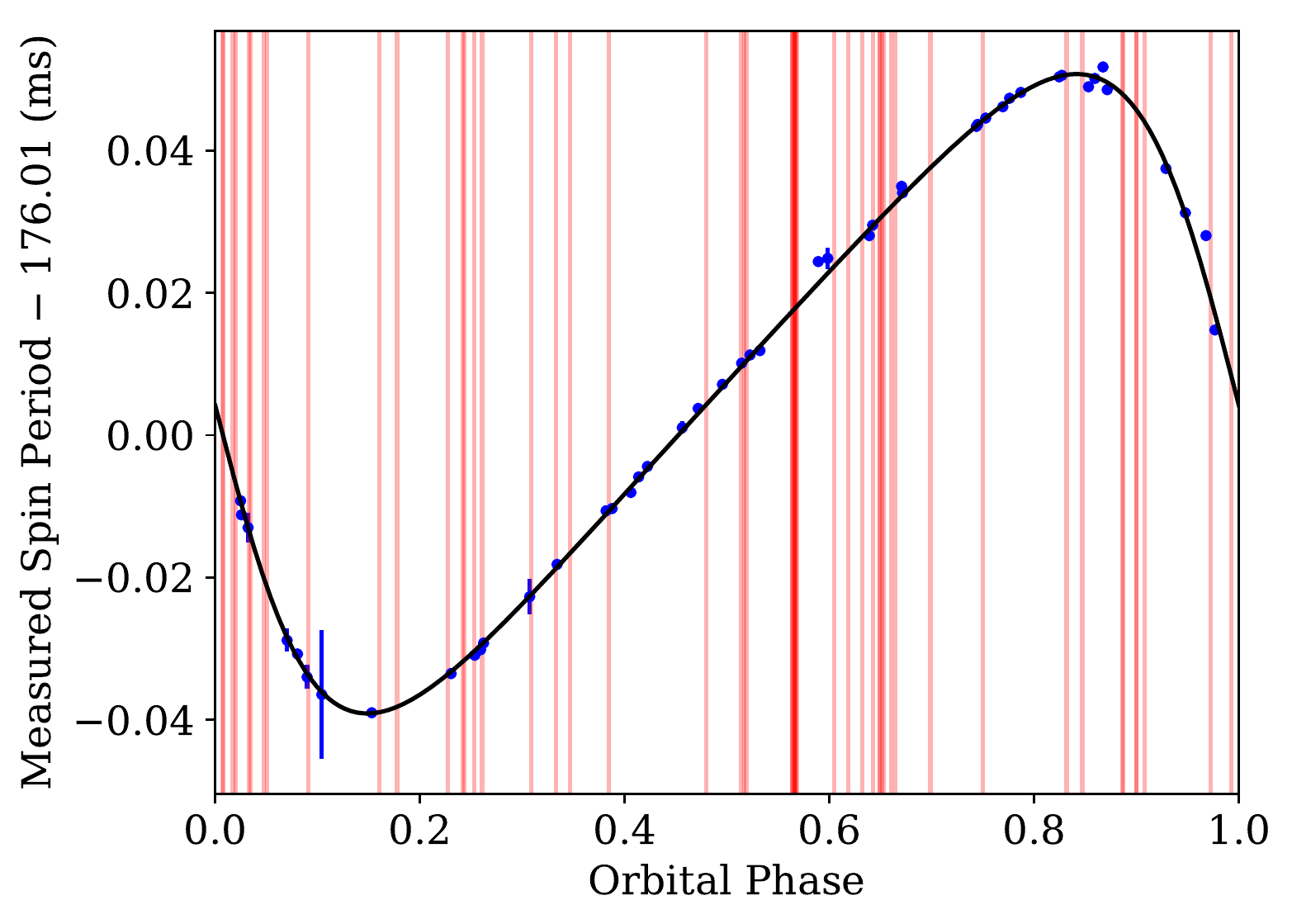}
    \caption{The observed Doppler-modulated $P$ of each epoch vs orbital phase. Red lines signify epochs of non-detection and do not appear to be associated with any particular phase of orbit. Overlaid in black is a calculated model of the variation of $P$ across the orbital phase. Orbital phases have been calculated in reference to the epoch of periastron.}
    \label{fig:binary_orbit}
\end{figure}

 From 2017--2019 we obtained  54 observations  of PSR~J1759+5036 on the GBT at 820 MHz with  40.96-$\rm \mu s $ time resolution and 2048 frequency channels. Initial timing efforts were complicated by the fact that the pulsar is in a binary system, and was also undetectable at 34 of the 54 epochs. We do not believe that extrinsic factors, such as RFI, were a significant contributor to the lack of detections. Since timing required initial estimates of the Keplerian binary parameters, we measured the barycentric $P$ at each epoch, and used methods from \citet{kep_param} to calculate rough approximations. These were then refined using \texttt{Tempo}\footnote{\url{https://sourceforge.net/projects/tempo/}} to build a phase-connected timing solution and to study the intermittent nature of the detections. We show in Figure~\ref{fig:binary_orbit} that there is no apparent connection between the binary orbital phase and the epochs of non-detections. This suggests that the lack of detections is either intrinsic to the pulsar emission mechanism or due to external interstellar medium effects, such as scintillation, possibly in combination with a poorly constrained position. A discussion of the scintillation will be presented in Section \ref{sec:intermit}.
 
 In May and June 2019 we conducted several observations with the  Karl G.\ Jansky Very Large Array (see Section 2.1), through which we determined a more accurate position which  was $7.22^{\prime} $ from our initial  position (i.e.\ that of PSR~J1800+5034). Given that the FWHM of the GBT 820-MHZ receiver is only $15^{\prime}$, our earlier observations suffered significantly from degraded sensitivity due to a reduced gain. The offset caused a $10\%$ and $44\%$ reduction in sensitivity for 350-MHz and 820-MHz observations, respectively.

We also carried out  a short, high-cadence timing campaign with the GBT at 350\,MHz. At this lower frequency, we had more success at detecting PSR~J1759+5036 than at 820\,MHz, with five of seven observations resulting in a detection.
This pulsar was also observed during 350-MHz GBNCC survey observations as a test source. We conducted another high-cadence campaign at 820\,MHz in December of 2019 that concluded in January of 2020. This gave us an additional 15 detections out of 24 observations.

\subsection{VLA Observations}

We observed PSR~J1759+5036 using the VLA during May and June 2019.  Since the source was previously observed
to be intermittent, we scheduled four separate observations to improve
our chances of detecting it.  These took place on MJDs 58614, 58615,
58652 and 58655.  Each session was 2.5\,hr long, with about 2\,
total on the pulsar.  During this time the VLA was in the
B configuration, with maximum baseline length of 11\,km.  We acquired
data using the VLA's interferometric pulsar binning mode, which averages
visibility data into a number of separate pulse phase bins following a
timing ephemeris. We used 32 pulse phase bins ($\sim$5.5~ms per bin) to record data spanning 1–2 GHz, with 1-MHz frequency resolution and 5-s dump time.

The data were processed as follows:  we used
\texttt{sdmpy}\footnote{\url{http://github.com/demorest/sdmpy}} to
re-align the data in pulse phase using the best available timing
solution, dedisperse at the known DM of the pulsar, and split the
original multi-bin data into separate data sets per bin as well as a
bin-averaged version.  All further calibration and imaging was done
using the Common Astronomy Software Applications package (CASA) \citep{casa} version 5.6.2.  The data were calibrated using
the standard VLA CASA calibration pipeline.  The flux scale was
referenced to 3C286 \citep{pb2017_flux}, and interferometric phases were
referenced to J1740+5211.  Calibration solutions were determined from
the bin-averaged data set and applied to each individual bin, which were
then imaged separately.  By subtracting the bin-averaged data from the
single-bin data, all confusing background sources are removed, resulting
in an unambiguous detection of the pulsar.

The pulsar was detected in all four observations, with flux density
ranging from 0.08 to 0.21\,mJy on different days. While this does show some epoch-to-epoch variations, the detectability at all four epochs suggests that scintillation at these higher frequencies is in the weak scattering regime, consistent with lack of structure in the  dynamic spectra.  
From the brightest
observation (MJD 58614) we determined a source position of RA = 17:59:45.66
$\pm$ 0.02s, Dec. = +50:36:57.05 $\pm$ 0.1\arcsec.  This observation had a
synthesized beamwidth of 4.9\arcsec\, by 4.1\arcsec\, at a position angle of
70\arcdeg.  The uncertainties quoted here reflect only the statistical
uncertainty due to noise in the data.  It is possible that systematic
effects may be present at the sub-beam level, however this position is
consistent with values obtained from the other three observations, as
well as with the later-determined timing position presented in
Table~\ref{tab:ephem}.

\section{Timing Analysis} 

\begin{figure}
    \centering
    \includegraphics[width=\linewidth]{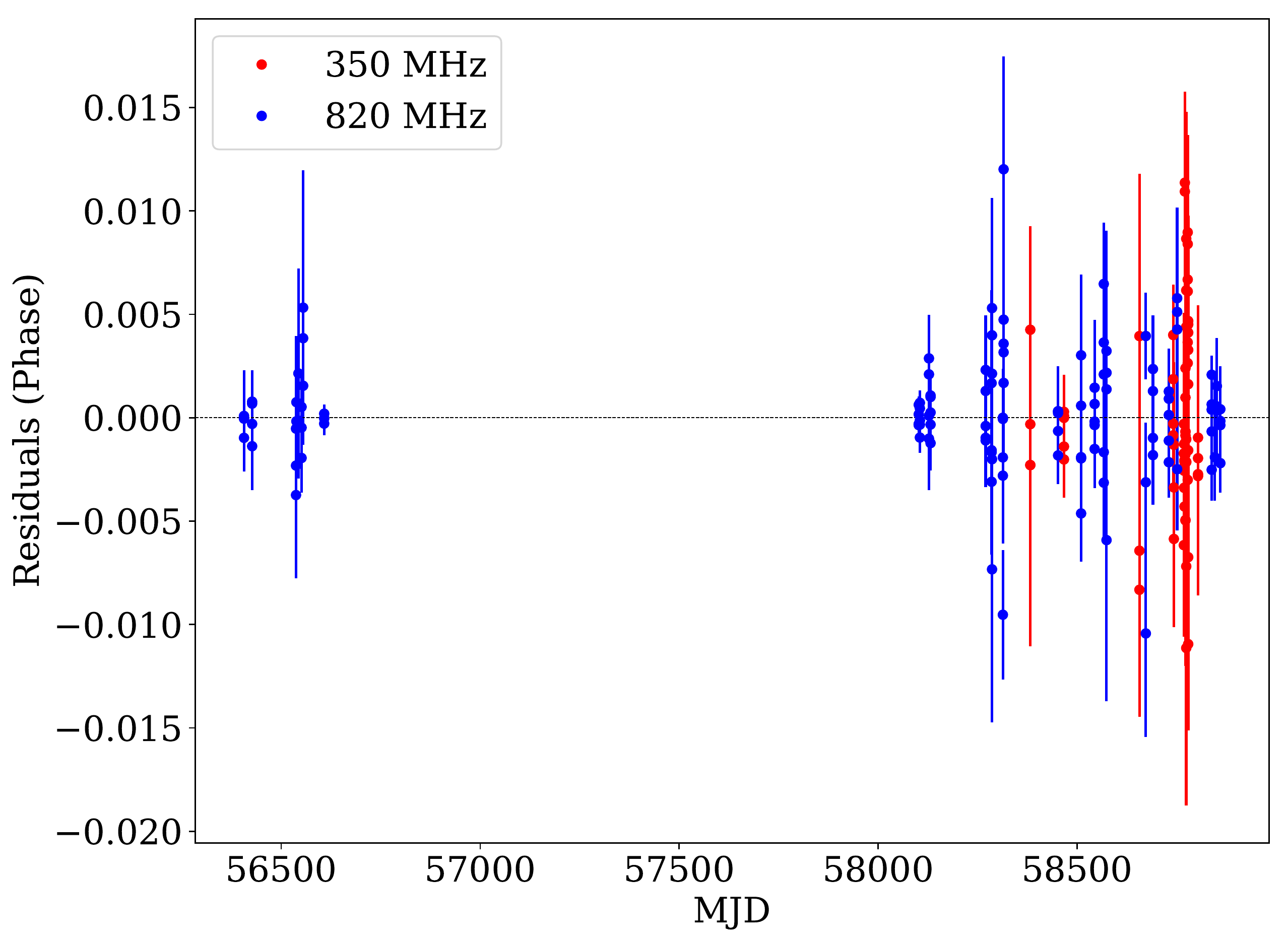}
    \caption{MJD vs post-fit residuals for 187 TOAs over 48 epochs. The TOAs from before MJD 58101 are from archived data taken from the PSR~J1800+5034 timing campaign. Red TOAs are calculated from 350-MHz data and blue TOAs are from 820-MHz data.}
    \label{fig:residuals}
\end{figure}

Our total data set spanned roughly seven years with a three year gap between archived timing data taken during 2013--2014 for PSR~J1800+5034 and the 2017--2020 data for PSR~J1759+5036. This can be seen in our timing residuals, shown in  Figure~\ref{fig:residuals}. Our ephemeris, shown in Table~\ref{tab:ephem}, is reported in TDB units, and uses the JPL DE436 solar system ephemeris and the DD binary model \citep{dd_model_1,dd_model_2}. We fit for DM by splitting five closely spaced individual observations taken at 350\,MHz into four frequency subbands and computing a single TOA from each subband. Our DM measurements indicate a distance of approximately 700\,pc using the NE2001 Galactic electron density model \citep{ne2001}, or 550\,pc using YMW16 model of Galactic electron density \citep{Yao_2017}. These distances are consistent within the errors for the two models.  

We were able to measure the Post-Keplerian (PK) parameter advance of periastron ($\dot{\omega}$) with high significance. Our reduced $\chi^{2}$ ($\chi_{\rm red}$) was initially 1.5 with 175 degrees of freedom (DOF), so we used a multiplicative error factor (EFAC) that accounts for random radiometer noise by applying a constant multiplier to all TOA error bars. Our applied EFAC was 1.25, which gave a $\chi_{\rm red}^{2}$ of 1.0. The root-mean-square (RMS) of our timing solution is 246\,$\rm \mu s$, or roughly 0.1\% of the pulsar spin period.

\begin{table}
  \begin{threeparttable}
    \caption{Timing solution for PSR~J1759+5036.}
    \label{tab:ephem}
     \begin{tabular*}{\linewidth}{l l}
        \toprule
        Measured Parameter & Value \\
        \midrule     
        Right Ascension (J2000) & 17:59:45.672(3)\\
        Declination (J2000) & +50:36:56.96(2)\\
        $P$ (s) & 0.17601634721733(8)\\
        $\dot{P}$ (s\,$\rm s^{-1}$) & $2.43(3)\times 10^{-19}$ \\
        Dispersion Measure ($\rm pc \; cm^{-3}$) & 7.775(3)\\
        \midrule
        Statistic and Model Parameters & \\
        \midrule
        Timing Data Span (MJD) & 56406--58859\\
        RMS Residual ($\rm \mu$s) & 246.54\\
        EFAC & 1.251 \\
        Number of TOAs & 187\\
        Binary Model & DD \\
        Solar System Ephemeris & DE436 \\
        Reference Epoch (MJD) & 57633\\
        \midrule
        Binary Parameters & \\
        \midrule
        Orbital Period, $P_{b}$ (days) & 2.04298385(3)\\
        Orbital Eccentricity, $e$ & 0.30827(12)\\
        Projected Semi-Major Axis (lt s) & 6.82461(3)\\
        Longitude of Periastron, $\omega$ (deg) & 92.142(2)\\
        Epoch of Periastron  (MJD) & 57633.09399(14)\\
        Advance of Periastron, $\dot{\omega}$ (deg $\rm yr^{-1}$) & 0.127(10)\\
        \midrule
        Derived Parameters & \\
        \midrule
        Surface Magnetic Field ($10^{9}$ Gauss) & 9.5 \\
        Spin-down Luminosity ($10^{30}$ erg\,$\rm s^{-1}$) & 9.0 \\
        Characteristic Age (Gyr) & 50 \\
        $\rm Dist_{DM}$ NE2001 (pc) & 711 \\
        $\rm Dist_{DM}$ YMW16 (pc) & 542 \\
        Mass Function & 0.081768(1)\\
        Minimum Companion Mass ($M_{\odot}$) & 0.7006\\
        Total Mass ($M_{\odot}$) & 2.62(3)\\
        $S_{350}$ (mJy) & 0.38(8) \\
        $S_{820}$ (mJy) &  0.11(2) \\
        $S_{1500}$ (mJy) & 0.12(2) \\
        \bottomrule
     \end{tabular*}
    \begin{tablenotes}
      \small
      \item Timing results are reported in units of TDB. In parentheses are the TEMPO-reported uncertainties in the last significant digit. The EFAC reported was used to achieve a reduced $\chi^{2}$ of 1.0. Position uncertainty is determined from the timing fit to position. The VLA observation position is consistent with timing position. We quote distances calculated with both the NE2001 \cite{ne2001} and YMW16 Galactic electron density models \cite{Yao_2017}.
    \end{tablenotes}
  \end{threeparttable}
\end{table}

We calculated the average flux density at 350 ($ S_{350}$) and 820 ($S_{820}$) MHz, as seen in Table~\ref{tab:ephem}, using the respective composite profiles integrated across pulse phase, using the radiometer equation \citep{1985Dewey}. For $S_{350}$, we assumed a bandwidth of 80\,MHz, a system temperature ($T_{\rm sys}$) of 46\,K, and a sky temperature ($T_{\rm sky}$) of 38\,K \citep{1982}. For $S_{820}$ we assumed a bandwidth of 200\,MHz, a $T_{\rm sys}$ of 29\,K, and a $T_{\rm sky}$ of 4.2\,K. $T_{\rm sky}$ values were calculated from the \cite{tsky} sky-maps and scaled to the appropriate frequency with a spectral index of --2.6. We calculated the off-pulse noise of each profile and then used the expected radiometer noise to scale the entire profile accordingly to determine the flux density. We estimate an approximate error of 20\% for each flux measurement made using the radiometer equation. From fitting a power law to the three points we get a spectral index of $-1.0 \pm 0.4$.

\subsection{Single-Pulse Analysis}

Using the \verb+PRESTO+ package \verb+single_pulse_search.py+ \citep{presto}, we searched for single pulses with S/N greater than 7 at every epoch between  dispersion measures of 0 and 20\,pc\,cm$^{-3}$ with  match-filtered box car widths ranging of 30 times the sample size. We determined whether pulses were indeed from PSR~J1759+5036 and not due to RFI by first requiring that the S/N be modeled as a Gaussian distribution with peak S/N at a DM between  7 and 8.25\,pc\,cm$^{-3}$, a range determined by the distribution of measured DMs on the detectable epochs. Through this method we found that $20\%$ of observations showed detectable single pulses. We did not observe any clear difference in the rate of single pulse detection for different observing frequencies.

\begin{figure}
    \centering
    \includegraphics[width=\linewidth]{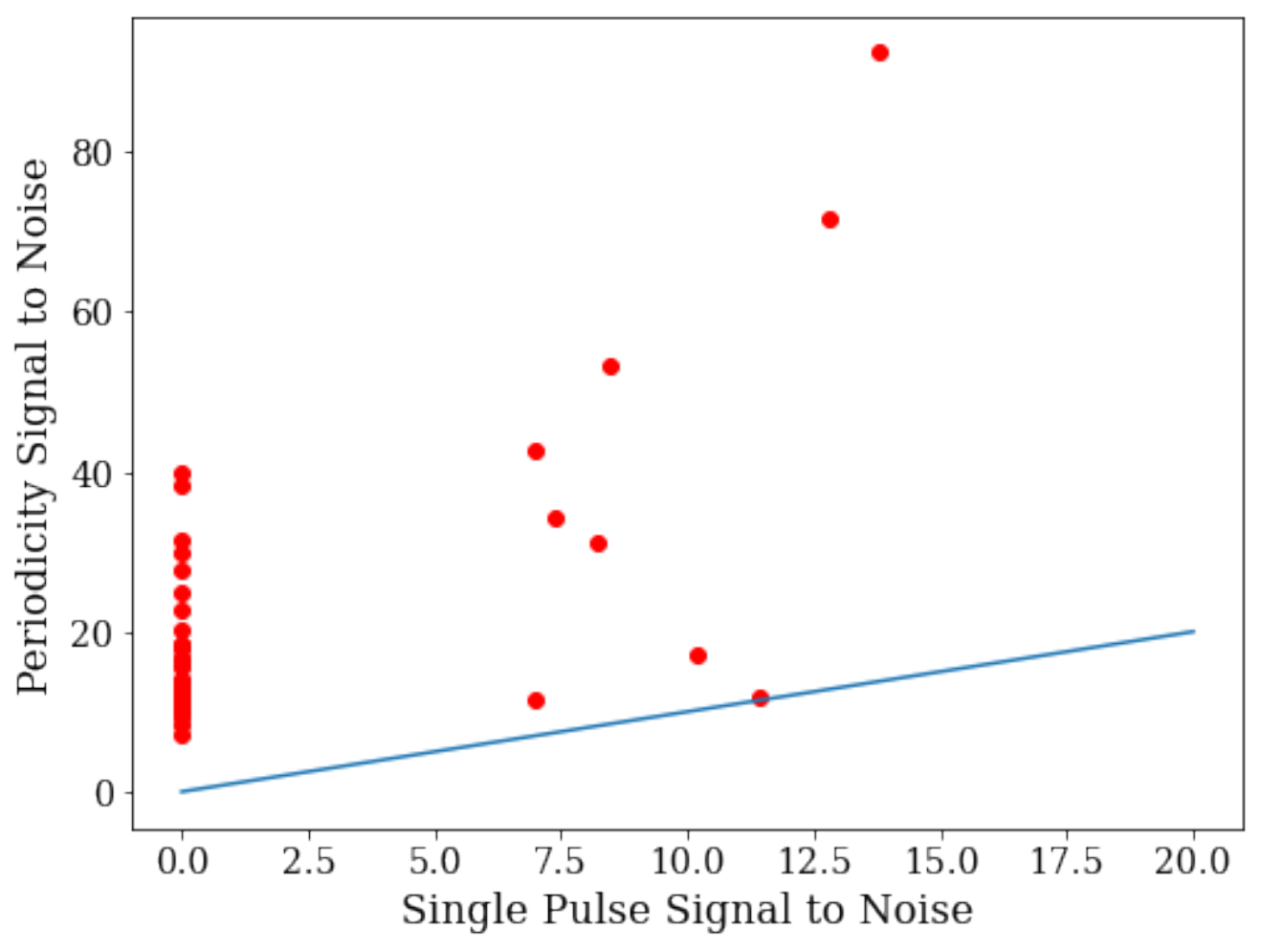}
    \caption{ Folded profile periodicity search S/N vs the S/N of the brightest detectable single pulse at each epoch of the timing observations,  with the line representing a slope of 1. On all days, the pulsar is equally or better detected in a periodicity search. Here, a value of zero indicates the pulse was not detectable using that method. The original detection is omitted in this plot.}
    \label{fig:periocityVs}
\end{figure}

Since this pulsar was originally detected only through its single pulses, we explored whether it is typically better detected in this way. In Figure~\ref{fig:periocityVs} we show the ratio of the S/N of the composite profile along with the S/N of the brightest single pulse. This shows that PSR~J1759+5036 is almost always detected at higher S/N values through periodicity searches than through single pulses, indicating that it is not an RRAT \citep{shapiro2018radio}. We ran the single-pulse search on the entire dataset, and only in the first search observation was the pulsar detected only through single pulses.

\begin{figure}
    \centering
    \includegraphics[width=\linewidth]{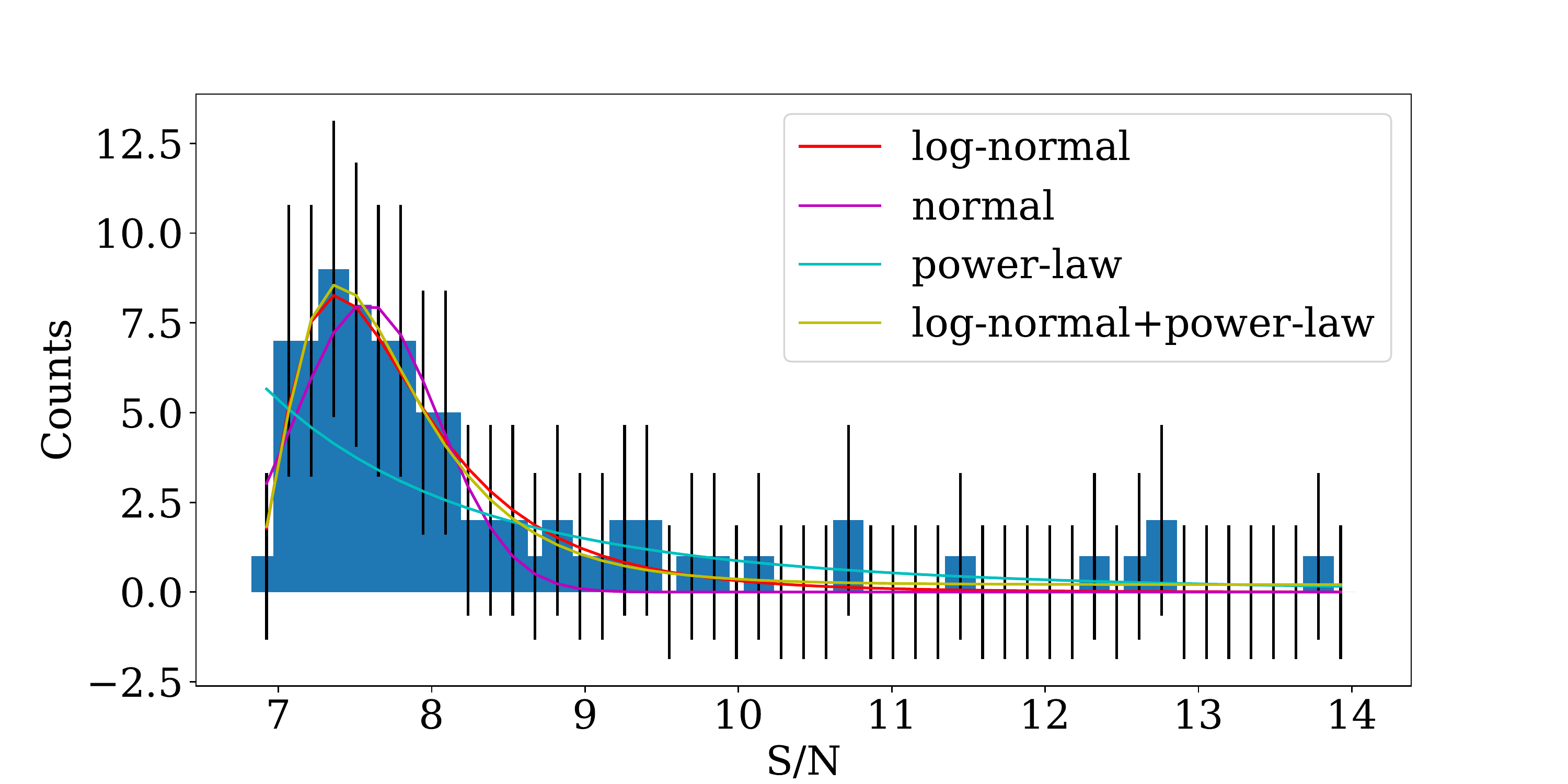}
    \caption{Histogram of the S/N of single pulses with 1-$\sigma$ error bars, using a minimum S/N cutoff of 7. Overlaid are the log-normal, normal, power-law, and log-normal plus power-law distributions that were fit to the data.}
    \label{fig:signalToNoise}
\end{figure}

We also wanted to see if the single pulse S/N distribution had a power-law tail, like those seen in RRATs and systems that emit giant pulses \citep{single_pulse_2018}. In Figure \ref{fig:signalToNoise}, we show a histogram of the detected single-pulse S/N values recorded at a central frequency of 820\,MHz. We show 1-$\sigma$ error bars as calculated in \cite{gehrels_1986}. To study the behavior of the S/N amplitudes we fit log-normal, normal, power-law, and combined log-normal/power-law distributions to the data. Given the large error bars on each bin, all four fits had  low chi-squared values, but  due to the low number of counts per bin we can not make any firm conclusions. With future data-sets we may be able to distinguish between these models and better understand how the single-pulse properties of this pulsar compare to those of other pulsars and RRATs.

\section{Discussion}

\begin{figure}
    \centering
    \includegraphics[width=\linewidth]{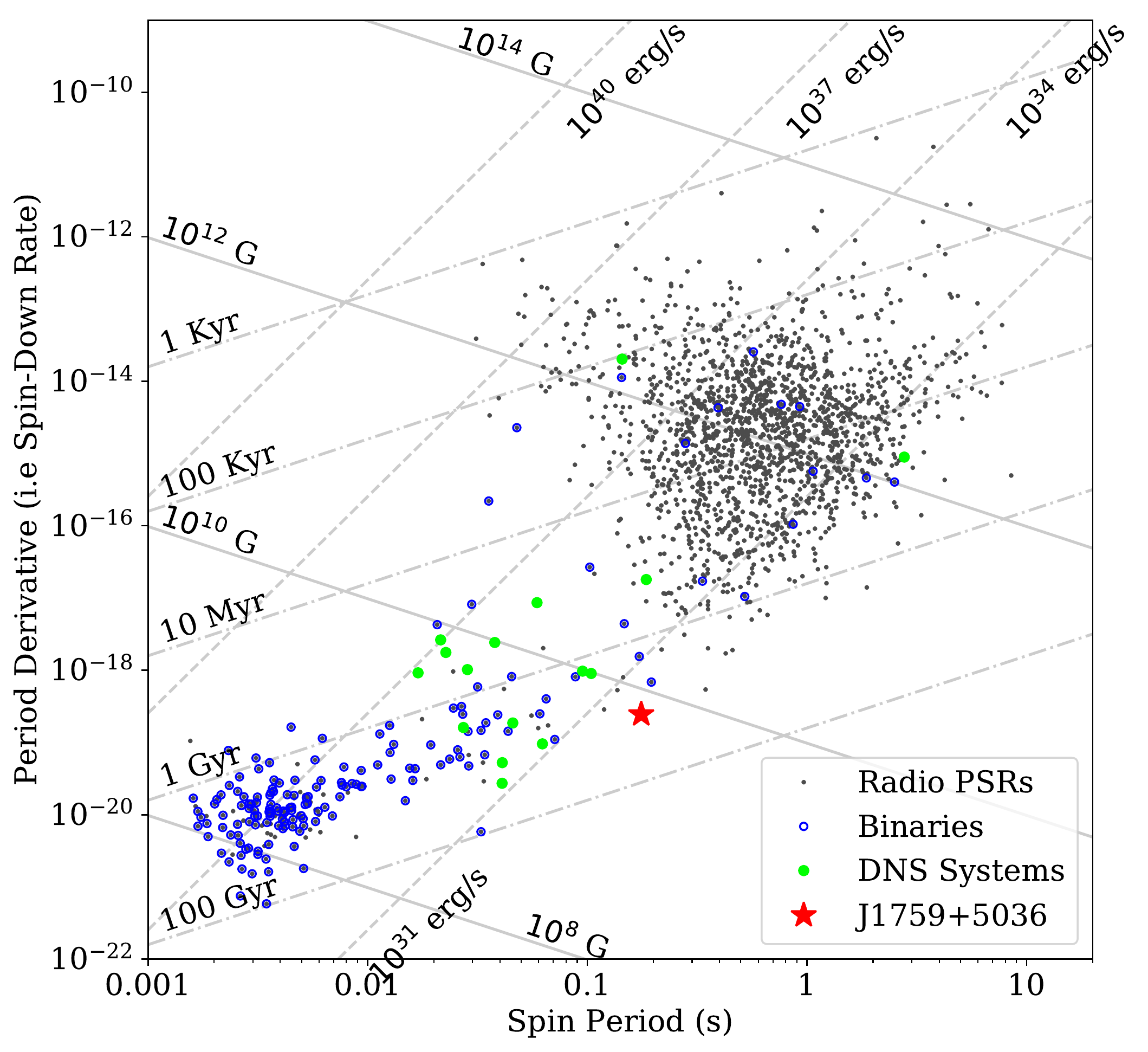}
    \caption{$P$ vs $\dot{P}$ of all known pulsars, with binary pulsars  in blue, DNS systems in green, and PSR~J1759+5036  in red. The parallel lines correspond to constant values of the characteristic age (dashed dotted), spin-down energy loss rate (dashed), and surface dipole inferred magnetic field (solid) (see e.g. \citealt{pulsar_handbook} for definitions). }
    
    \label{fig:ppdot}
\end{figure}

The small $\dot{P}$ and relatively large $P$ place
PSR~J1759+5036 in a sparsely populated region of $P$ and $\dot{P}$ space (see Figure~\ref{fig:ppdot}). The relatively low magnetic field of about  $10^{10}$ G suggests that this pulsar is partially recycled, but some process halted accretion from the companion before spin-up to very short periods. The observed high eccentricity is likely an artifact of the companion supernova explosion, and is consistant with those seen in DNS systems \citep{Tauris_2017,DNS_merger_rate}.

The time to merger due to gravitational wave radiation is  182\,Gyr, much longer than a Hubble time, meaning that PSR~J1759+5036's discovery will not impact DNS merger rate estimates \citep{binary_living_review,DNS_merger_rate}.

\subsection{Nature of the Binary System}

\begin{figure}
    \centering
    \includegraphics[width=\linewidth]{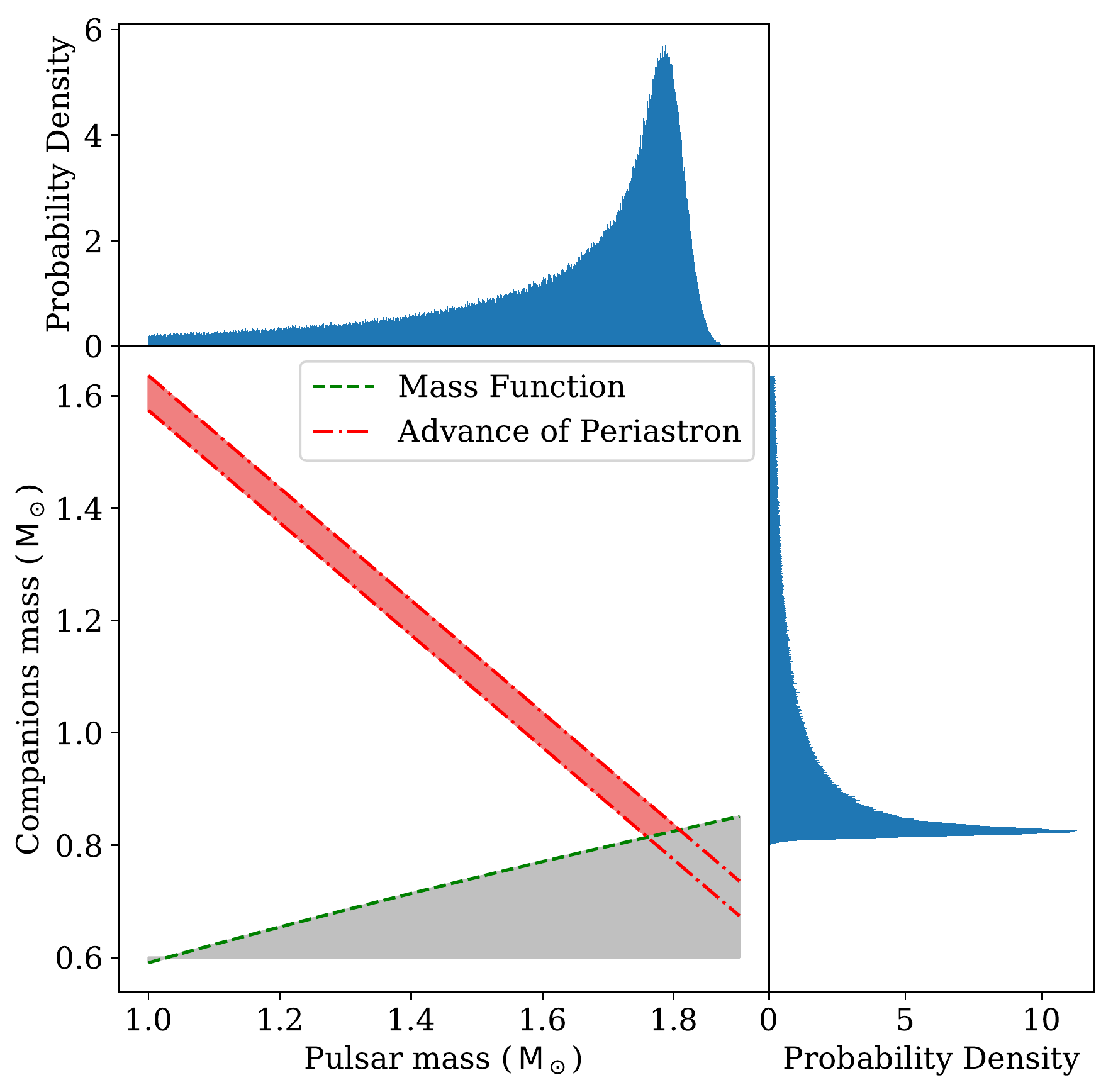}
    \caption{Allowed values of pulsar mass vs companion mass given the rate of advance of periastron (between the red dashed lines) and the mass function (above the green dotted line). The red shaded region indicates possible values of pulsar mass and companion mass. Above and to the right are the probability distributions for the pulsar and companion mass respectively.}
    \label{fig:mass_mass}
\end{figure}

The mass function implies a minimum companion mass of 0.7\,$M_{\odot}$, assuming a system inclination of $90\degree$ \citep{binary_living_review}. Our measurement of $\dot{\omega}$ corresponds to a total system mass of 2.62 $\pm$ 0.03\,$M_{\odot}$. In Figure~\ref{fig:mass_mass} we have overlaid the possible pulsar and companion masses allowed by $\rm \dot{\omega}$ with those forbidden by the mass function. This allowed us to determine the maximum pulsar mass to be 1.8\,$M_{\odot}$. The probability distributions for pulsar and companion mass in the figure have been calculated using the measured $\dot{\omega}$ assuming random orbital inclinations. The 2$\sigma$ mass ranges are 1.26--1.79\,$M_{\odot}$ for the pulsar and 0.84--1.37\,$M_{\odot}$ for the companion, with a 0.16 probability that the companion mass is greater than 1.2\,$M_{\odot}$. If the companion is indeed a NS, as seems likely given its spindown and orbital properties, the inclination angle must be less than 44\degree (assuming the NS mass must be greater than 1.2\,$M_{\odot}$)\citep{thorsett_ns_mass}.

\begin{figure}
    \centering
    \includegraphics[width=\linewidth]{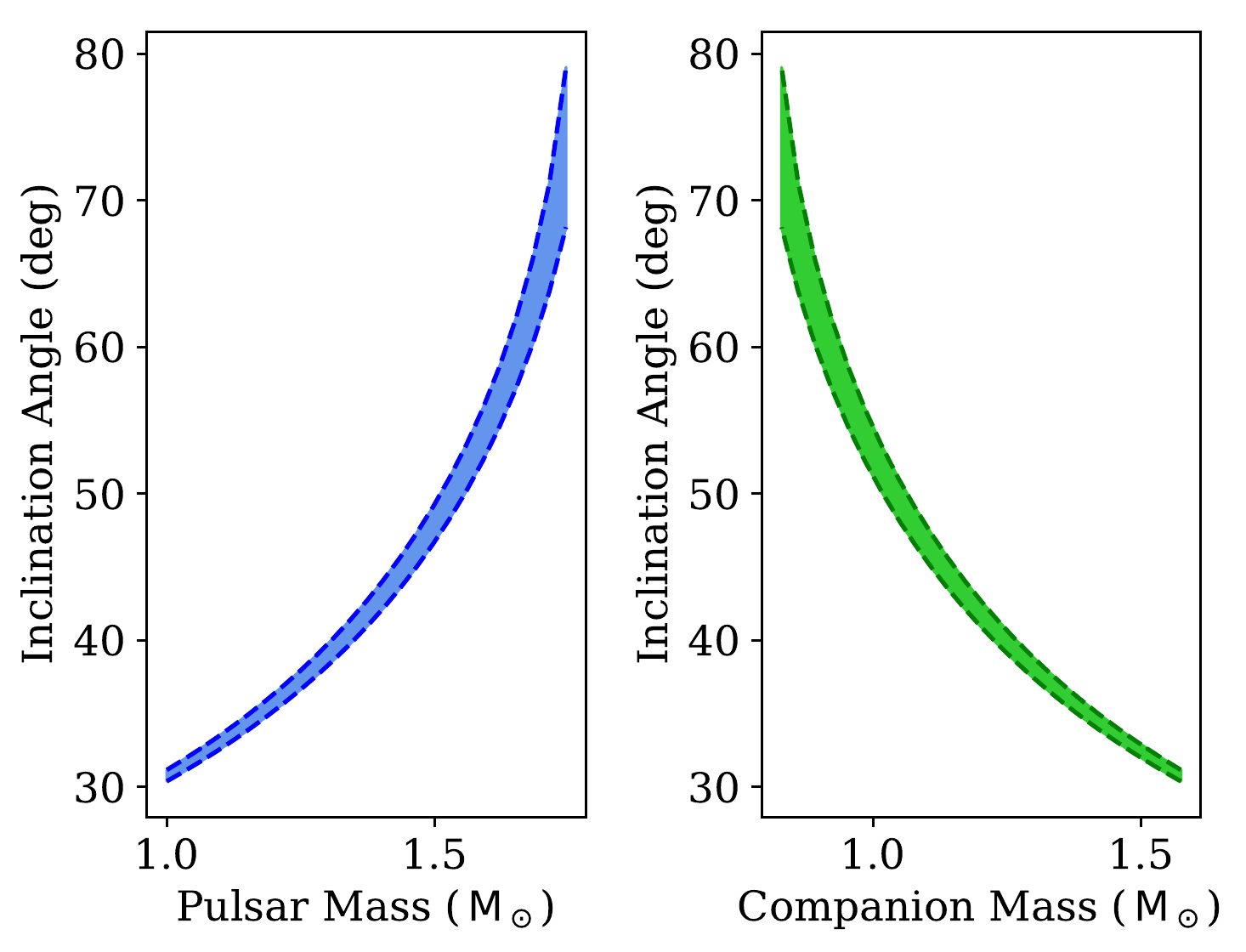}
    \caption{Left:  Potential pulsar masses vs  projected inclination angle implied by each  mass as shown by the shaded region. Right: Potential masses of the companion mass vs inclination angle, with the shaded region indicating possible values.}
    \label{fig:inclination}
\end{figure}

In Figure~\ref{fig:inclination} we show projected values for the inclination angle of the binary system using the range of possible pulsar and companion masses determined by the total system mass and the mass function. We can infer from this that the inclination angle must be between  35$\degree$ and 75$\degree$.

\subsection{Optical Constraints on a Companion}
\label{sec:companion}
\begin{figure}
    \centering
    \includegraphics[width=\linewidth]{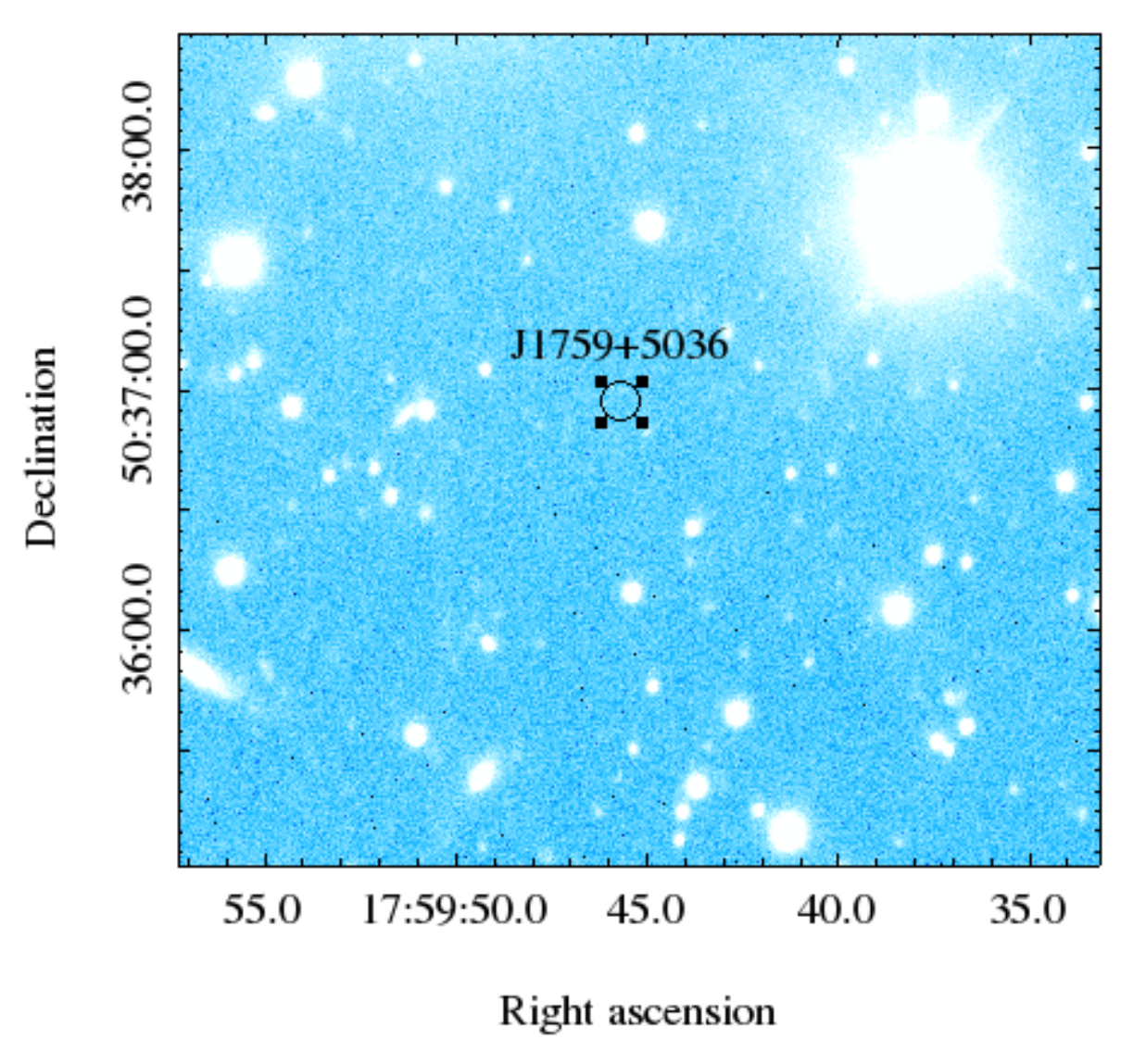}
    \caption{Image from the LCO observation to look for an optical counterpart for PSR~J1759+5036. The black circle represents a $5^{\prime\prime}$ diameter region around the pulsar, much larger than the  $<1^{\prime\prime}$ error in position.}
    \label{fig:DSS}
\end{figure}

We conducted optical observations with the Las Cumbres Observatory 2-meter telescope on Haleakala on 30 Apr 2020. Three consecutive 500\,s exposures were taken with the SDSS $ r^{\prime}$ filter (median  MJD 58969.5637447) and then another three 500s exposures with the SDSS $g^{\prime}$ filter (median MJD 58969.5457716). These were then median added for each filter and aperture photometry was attempted on them. We used images whose dark subtraction, bias correction, flat-fielding
and plate solutions were generated by the LCO {\it Banzai} pipeline. In neither filter was a counterpart detected at the position of the pulsar (Figure~\ref{fig:DSS}). By measuring the faintest nearby stars using two SDSS catalog (release 12) stars as references, we estimate a limiting magnitude of $\sim$23.5 in both filters at the position of PSR~J1759+5036.  

\begin{figure}
    \centering
    \includegraphics[width=\linewidth]{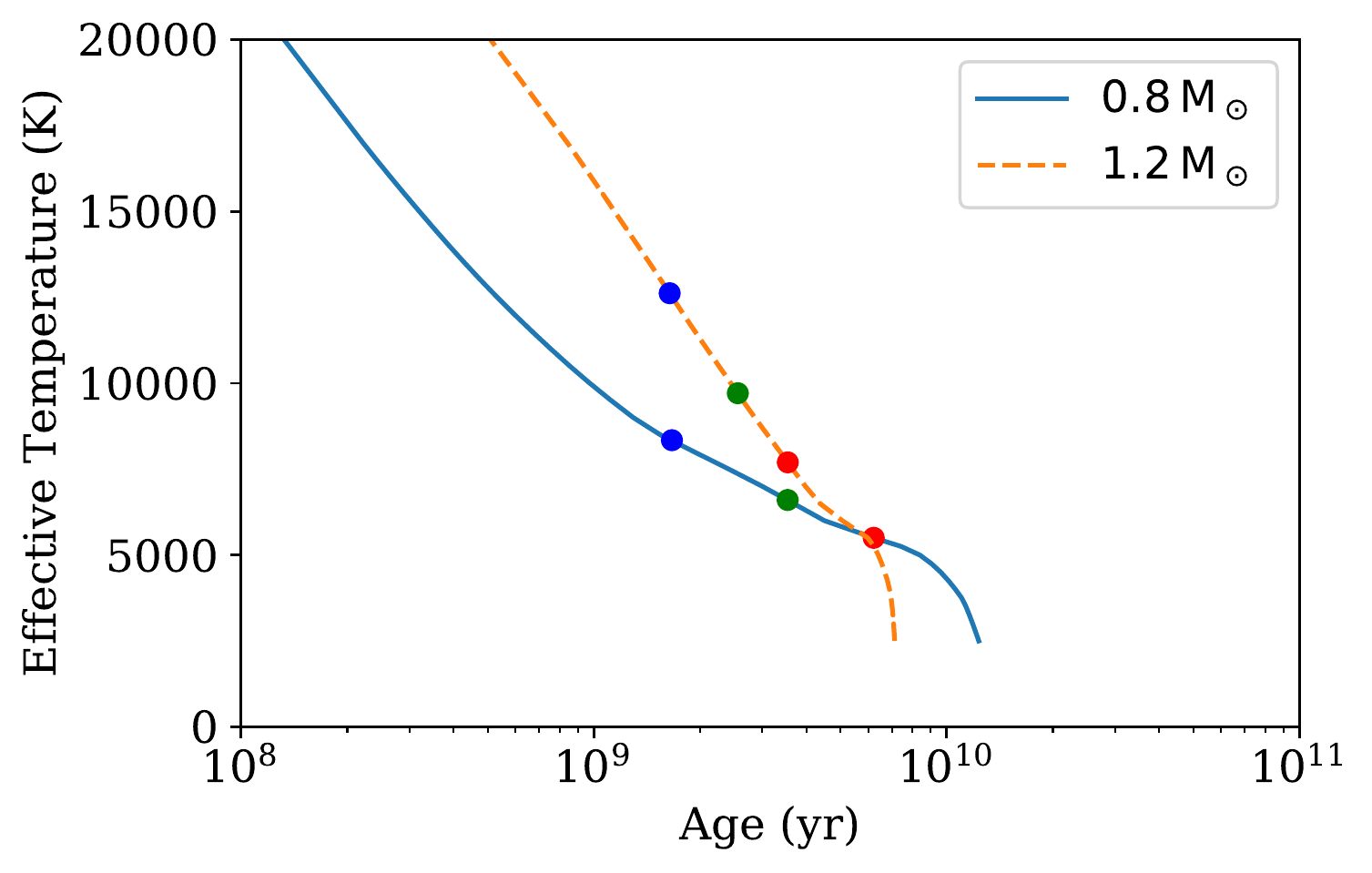}
    \caption{Cooling curves for 0.8\,$M_{\odot}$(blue) and 1.2\,$M_{\odot}$(orange) WD stars with effective temperature upper limits of  and age lower limits indicated for distances of 500 pc (red), 700 pc (green), and 1000 pc (blue). For an 0.8\,$M_{\odot}$ WD we have limits of $<5500$\,K and $>6.2$\,Gyr at 500\,pc, $<6600$\,K and $>3.5$\,Gyr at 700\,pc, and $<8400$\,K and $>6.2$\,Gyr at 500\,pc. For a 1.2\,$M_{\odot}$ WD we have limits of $<7700$\,K and $>3.5$\,Gyr at 500\,pc, $<9700$\,K and $>2.6$\,Gyr at 700\,pc, and $<12600$\,K and $>1.6$\,Gyr at 500 pc. }
    \label{fig:wd_cooling}
\end{figure}

Using 23.5 as an apparent magnitude minimum limit for a possible WD companion, we determined temperature and age limits for distances of 500, 700, and 1000\,pc to account for the uncertainty in the DM distance. 
In Figure \ref{fig:wd_cooling}, we have plotted WD cooling curves\footnote{\url{http://www.astro.umontreal.ca/~bergeron/CoolingModels/}}, with the limits for each distance indicated \citep{cool_curves_orig,cool_curves_1,cool_curves_2,cool_curves_3,wd_cooling_curves,cool_curves_4,cool_curves_5}. For this analysis we assumed a hydrogen WD atmosphere, used dustmaps to take extinction into account, and used cooling curves for masses of 0.8\,$M_{\odot}$ and 1.2\,$M_{\odot}$, which are the lower and upper mass limits for a WD companion for this system \citep{wd_cooling_curves}. 
For a 0.8\,$M_{\odot}$ companion at a distance of 700\,pc the effective temperature is $<$ 6600\,K and the age is $>$ 3.5\,Gyr, which is based on the more constraining $g^{\prime}$ limit. At the same distance, for a 1.2\,M$_{\odot}$ companion, the effective temperature $<9700$\,K and the age $>2.6$\,Gyr. 
These limits are not particularly constraining as, if the companion has nearly the same age as the pulsar, it would have a temperature below either of these limits (see Figure~\ref{fig:wd_cooling}) and likely be too faint to be detected even with deeper optical observations.

Being unable to detect an optical counterpart at the position of PSR~J1759+5036, we have ruled out a main sequence star companion (already very unlikely due to the tight orbit) and most white dwarf (WD) stars (Figure~\ref{fig:DSS}). Any main sequence star of comparable luminosity and size to the Sun at our DM estimated distances (Table~\ref{tab:ephem}) would have a magnitude of 13--15 and thus would be easily visible. Optical observations at a magnitude of 26--27 would be needed to confidently rule out or confirm a WD companion.

While most pulsar-WD binaries are circular, there are several systems with eccentric companions, such as PSR~J2305+4707 and PSR~J1755$-$2550. However, these binaries have some significant differences in properties from PSR~J1759+5036. Both have limited evidence of recycling and have significantly lower characteristic ages of 29.7\,Myr and 2.1\,Myr, respectively, whereas PSR~J1759+5036 shows evidence of partial recycling and a much larger  characteristic age  \citep{compact_companion,eccentric_binary}. Therefore, we argue it is most likely a DNS system.

\begin{figure}
    \centering
    \includegraphics[width=\linewidth]{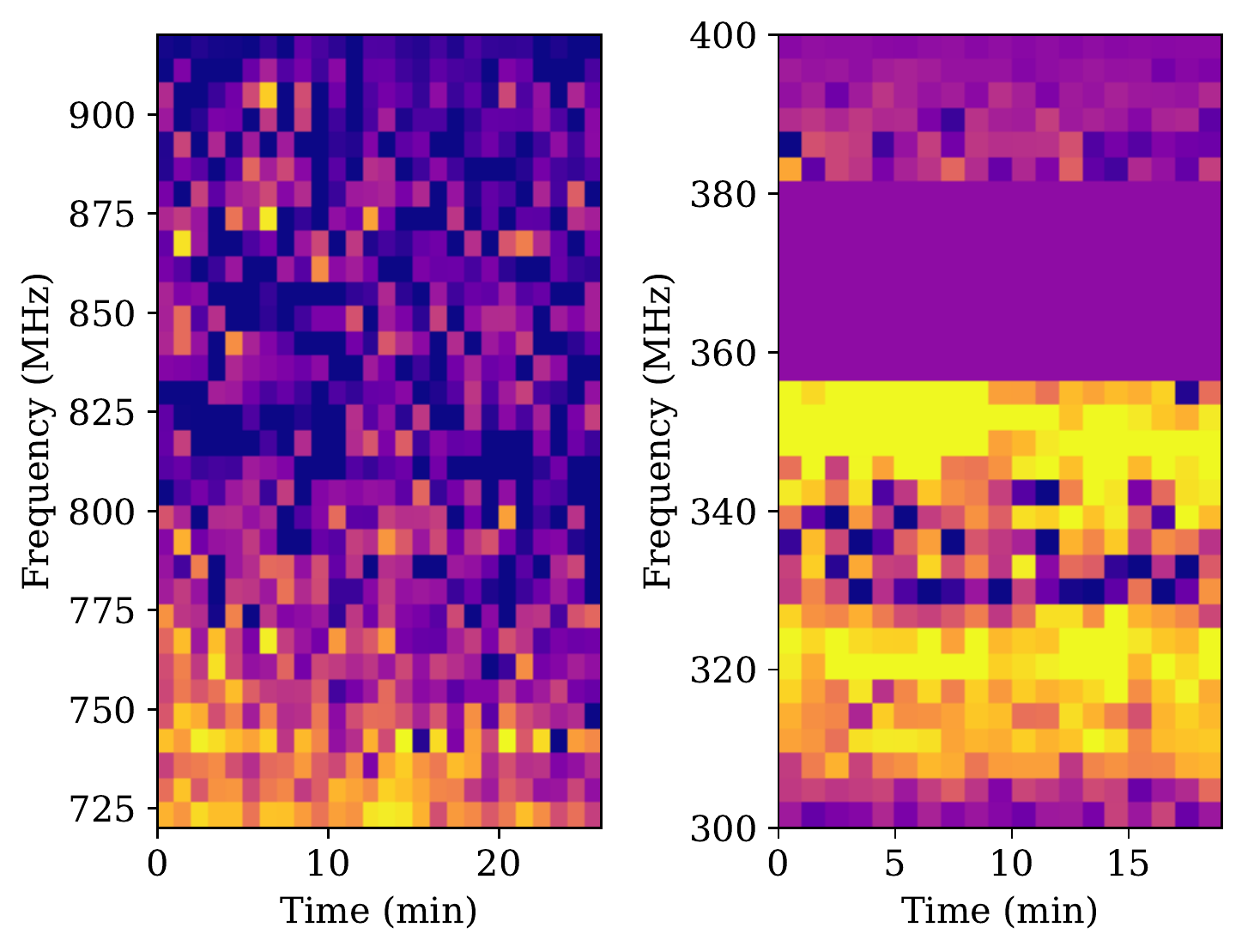}
    \caption{Left: A dynamic spectrum on epoch 58131 at 820\,MHz. Intensity is plotted against frequency bins of roughly 6\,MHz, and time bins of 60\,s. Right: Dynamic spectrum on epoch 58466 at 350\,MHz. On-pulse intensity is plotted vs frequency and time, with  time bins of 60\,s and frequency bins of roughly 3\,MHz. The purple strip between 360 and 380\,MHZ is a region of bright RFI that was removed from the data.}
    \label{fig:dynam}
\end{figure}

\subsection{Intermittency}
\label{sec:intermit}
The low DM of PSR~J1759+5036 suggests that diffractive interstellar scintillation may play a role in the intermittency we observed, shown in Figure~\ref{fig:binary_orbit}. To study this, we created dynamic spectra of each detected epoch by first folding the data in time sub-integrations of 60~s using the \verb+fold_psrfits+ package from \verb+PSRCHIVE+ \citep{psrchive}. Then, we used \verb+pam+ to fold the data into frequency sub-integrations of 3\,MHz and plotted the intensity of each frequency/time bin and corrected the bandpass for instrument variation using the python package \verb+PyPulse+ \citep{pypulse}. In Figure~\ref{fig:dynam} we have shown dynamic spectra from two epochs where scintillation was observed.

The NE2001 electron density model predicts a scintillation bandwidth of $\sim$26\,MHz at 350\,MHz and $\sim$40\,MHz at 820\,MHz at the position and DM of J1759+5036 \citep{ne2001}. 

Given the measured timing residual and the 6-year baseline of our timing data, we can place an upper limit on the transverse velocity of 16\,$\rm km\,s^{-1}$. This is rather low, but consistent with other DNS systems such as PSR~J0737-3039A/B ($\sim$9\,$\rm km\,s^{-1}$) and PSR~B1913+16 ($\sim$22\,$\rm km\,s^{-1}$) \citep{double_pulsar_velocity,B1913_velocity}. For this velocity we estimate a scintillation timescale of $\sim$6400\,s at 350\,MHz and $\sim$9300\,s at 820\,MHz. Since these timescales are significantly longer than any of our observations this is consistent with our inability to resolve scintles in time.

On epoch 58466, we resolved two scintles in frequency with the more visible scintle having a bandwidth of roughly 25\,MHz, indicating that the NE2001 prediction is accurate (Figure~\ref{fig:dynam}). The second scintle is not fully resolved due to  RFI excision. For the 820\,MHz data we were unable to fully resolve scintles on any of the epochs, but on one epoch, MJD 58131, we detected  a partially resolved scintle that is cut off at the edge of the receiver band (Figure~\ref{fig:dynam}). It is easy to see that this epoch may have resulted in a non-detection if this scintle was centered at a slightly lower frequency. Thus, scintillation may explain the numerous non-detections of this pulsar at 820\,MHz.

On one epoch, MJD 58772, PSR~J1759+5036 was only detected in the second and third of three contiguous observations. The first observation was six minutes, the second 30 minutes, and the third  five minutes. There was a two minute gap between the first two observations and a ten minute gap between the second and third observation. The first detection was strong, with a S/N of 17.2, whereas the second had a S/N of 9.6. These timescales are consistent with those expected due to scintillation, and is further evidence that scintillation may be responsible for the non-detections.

\begin{figure}
    \centering
    \includegraphics[width=\linewidth]{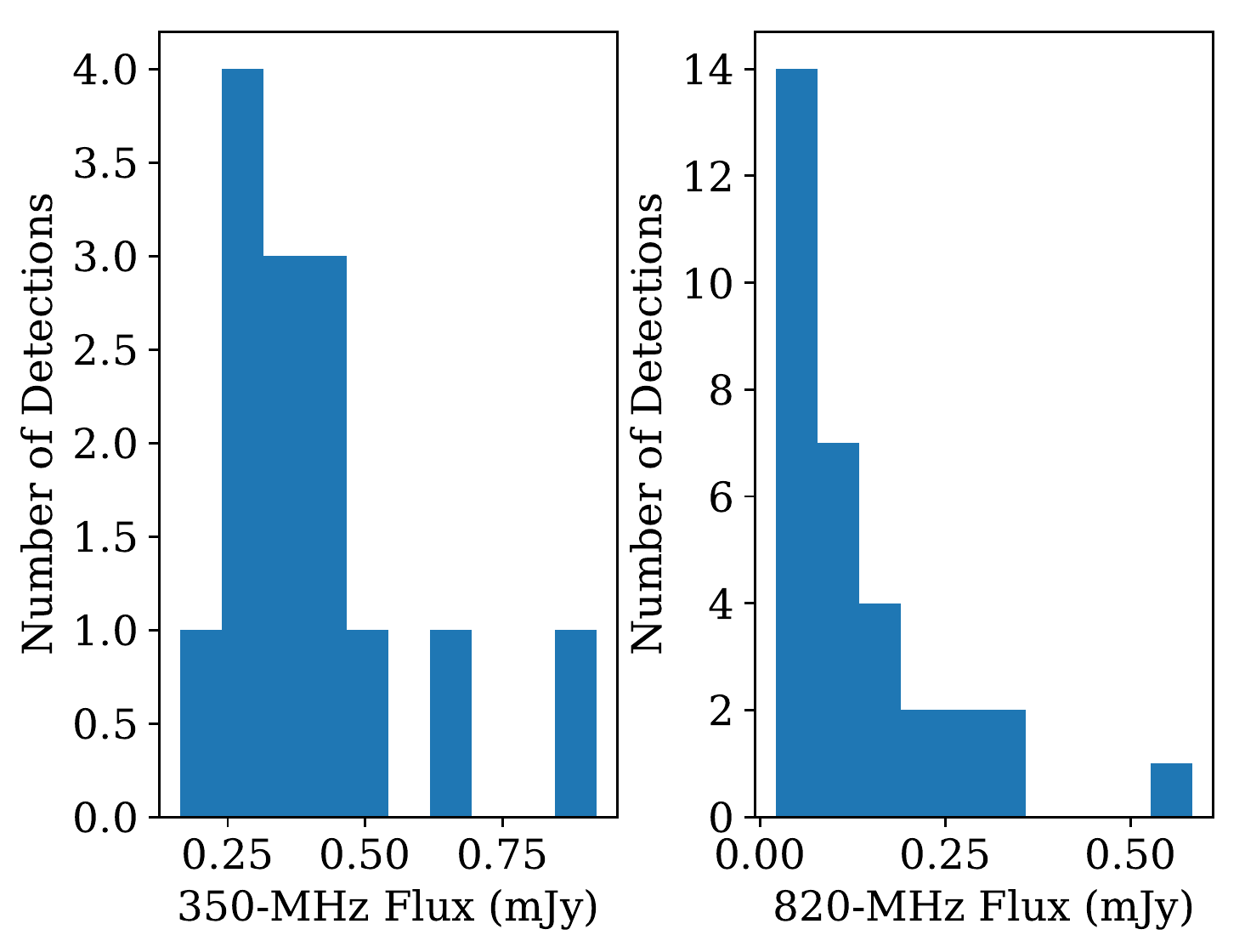}
    \caption{The distribution of flux densities measured at each epoch at 350\,MHz (left) and 820\,MHz (right). Flux densities that were calculated from observations impacted by the position offset were corrected to reflect the degraded sensitivity.}
    \label{fig:flux}
\end{figure}

In Figure~\ref{fig:flux}, we show histograms of the flux variations at both 350 and 820\,MHz. 
The 820-MHz distribution shows the exponential tail expected for variability due to diffractive interstellar scintillation \citep{scheuer,hesse}. Flux densities that would have been impacted by the position offset in our earlier observations have been corrected to reflect the degraded sensitivity.

\begin{figure*}
    \centering
    \includegraphics[width=\textwidth]{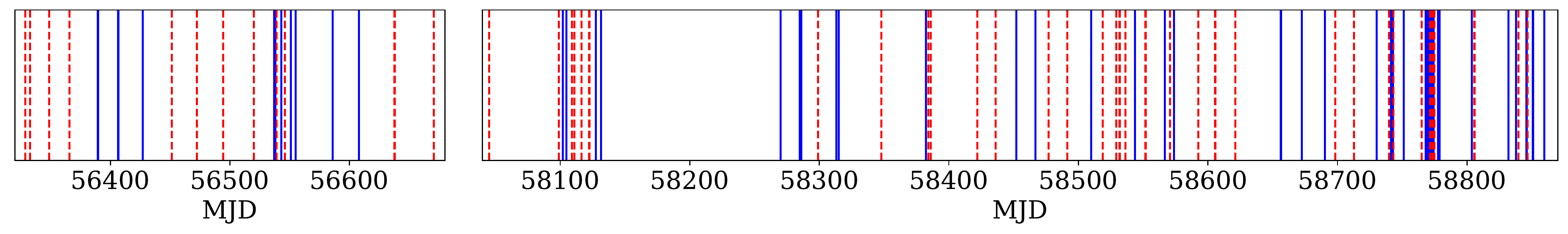}
    \caption{Epochs of observation on which PSR~J1759+5036 was detected (blue sold lines) and not detected (red dotted lines). Left: 2013--2014 dataset. Right:  2017--2020 dataset.}
    \label{fig:on_off}
\end{figure*}

We ran a Lomb-Scargle analysis (e.g., \citealt{scargle_orig,Lombscargle}) to determine if there was any periodicity to the on and off periods of PSR~J1759+5036, shown in Figure~\ref{fig:on_off}. The analysis tested 122 trial periods ranging from 50--2530 days, and the most significant signal had a periodicity  of 171 days with a false alarm probability of 0.05. This is only marginally significant; more data are needed to determine whether a true periodicity might exist.  

\section{Conclusions and Future Work}

In this paper we report on the discovery and timing campaign  for PSR~J1759+5036, a 176-ms binary pulsar with a 2.04-day eccentric orbit about a likely neutron star companion. Over our seven-year dataset, this pulsar has been highly intermittent with an approximately 45\% detection rate which we believe to be largely attributed to scintillation, possibly combined with  a poorly constrained position that we then determined through observations taken with the VLA.

We were able to measure all five Keplerian parameters as well as a single PK parameter, $\dot{\omega}$. This allowed us to determine the minimum companion mass and total system mass, which we used to place limits on possible pulsar and companion masses as well as system inclination angle. Due to the eccentricity (0.308) of the system and lack of optical counterpart, we believe the candidate is most likely a neutron star. We have ruled out a main sequence companion and most white dwarf stars (see Section~\ref{sec:companion}).

Continued follow-up observations and timing of PSR~J1759+5036 are expected to eventually produce a measurement of a second PK parameter, which would allow us to fully constrain the pulsar and companion masses. 
Using simulated TOAs, we estimate that approximately 12 years of data would be needed to measure $\gamma$, the time dilation and gravitational redshift parameter, which would also determine the inclination angle of the system. We also hope to be able to measure the proper motion, which would allow us to calculate more accurate estimates of distance and pulsar velocity. Currently, follow-up observations are being conducted on the Canadian Hydrogen Intensity Mapping Experiment telescope with weekly cadence, the results from which are expected to be included in future publications.

\acknowledgements
MAM, JKS, HB, MD, PBD, WF, EF, DLK, RSL, AEM, SMR, AS, CS, XS, IHS, and MS are members of the NANOGrav Physics Frontiers Center, supported by NSF award number 1430284. VMK holds the Lorne Trottier Chair in Astrophysics \& Cosmology, a Distinguished James McGill Chair, and receives support from an NSERC Discovery Grant, the Herzberg Award, CIFAR, and from the FRQ-QNT Centre de Recherche en Astrophysique du Quebec. MAM, GYA, and MGM are also supported by NSF award number 1458952. We thank the WVU Research Corporation for their purchase of observing time on the GBT which has supported some of the observations for this project. Pulsar work at UBC is supported by an NSERC Discovery Grant and by the Canadian Institute for Advanced Research.
The National Radio Astronomy Observatory and
Green Bank Observatory are facilities of the National Science Foundation
operated under cooperative agreement by Associated Universities, Inc. Computations were
made on the supercomputer Guillimin at McGill University\footnote{\url{ www.hpc.mcgill.ca}}, managed by Calcul Quebec and Compute Canada. The operation of this supercomputer is funded by the Canada Foundation for Innovation (CFI), NanoQuebec, RMGA and the Fonds de recherche du Quebec - Nature et technologies (FRQ-NT). 
VLA observations were taken as project 19A-387. GBT observations were taken under projects AGBT13A-458, AGBT17B-285, AGBT17B-423, AGBT17B-292, AGBT18A-482,  AGBT17B-325, AGBT18B-335, AGBT18B-360, AGBT19A-180, AGBT19A-486, AGBT19B-306, AGBT19B-327, and AGBT19B-320.

\textit{Software}: \verb+Astropy+\citep{astropy}, \verb+PRESTO+ \citep{presto}, \verb+PSRCHIVE+\citep{psrchive}, \verb+PINT+\citep{pint}, \verb+NumPy+\citep{numpy}, \verb+PyPulse+\citep{pypulse},\verb+TEMPO+,\verb+SciPy+, \verb+CASA+\citep{casa}, \verb+DS9+\citep{ds9}, \verb+sdmpy+ (\url{http://github.com/demorest/sdmpy})

\textit{Facilities}: Robert C. Bryd Green Bank Telescope (GBT), Karl G. Jansky Very Large Array (VLA), Las Cumbres Observatory (LCO)

\bibliographystyle{aasjournal}
\bibliography{manuscript.bib}
\end{document}